\documentclass[12pt]{article}
\usepackage{geometry}
\usepackage{graphicx} % Required for inserting images
\usepackage{float}
\usepackage{nameref}  % For naming references by section title
\usepackage{amsmath}
\usepackage{authblk}
\usepackage{comment}
\usepackage{xcolor}
\usepackage{url}

\usepackage{rotating}  % For sideways table
\usepackage{booktabs}  % For professional table formatting
\usepackage{multirow}  % For better table alignment

\title{Spectral Analysis of Light Interstitial Segregation Energies in Ni: The Role of Local Cr Coordination for Boron and Carbon}

\author[1,2]{Tyler D. Dole\v{z}al\footnote{corresponding authors: tyler.dolezal.1@us.af.mil; rodrigof@mit.edu; liju@mit.edu}}
\author[1]{Rodrigo Freitas$^*$}
\author[1,3]{Ju Li$^*$}
\affil[1]{Department of Materials Science and Engineering, Massachusetts Institute of Technology, Cambridge, MA, USA}
\affil[2]{Department of Engineering Physics, Air Force Institute of Technology, Wright-Patterson Air Force Base, OH, USA}
\affil[3]{Department of Nuclear Science and Engineering, Massachusetts Institute of Technology, Cambridge, MA, USA}

\begin{document}

\maketitle

\begin{abstract}
\noindent
Understanding interstitial segregation in chemically complex alloys requires accounting for chemical and structural heterogeneity of interfaces, motivating approaches that move beyond scalar descriptors to capture the full spatial and compositional spectra of segregation behavior. Here, we introduce a spectral segregation framework that maps distributions of segregation energies for light interstitials in Ni as a function of local Cr coordination. Boron exhibits a broad, rugged energy spectrum with significant positional flexibility whereas carbon remains confined to a narrow spectrum with minimal displacement. At the free surface, Cr-rich coordination destabilizes both interstitials (e.g., positive segregation energies), in sharp contrast to the stabilizing role of Cr at the GB. This inversion establishes a natural segregation gradient that drives interstitials away from undercoordinated internal surfaces and toward GBs. These results underscore the limitations of single-valued segregation descriptors and demonstrate how a distributional approach reveals the mechanistic origins of interstitial--interface interactions in chemically heterogeneous alloys.
\end{abstract}

The segregation of light interstitials such as boron and carbon to internal interfaces critically influences the thermomechanical performance of structural alloys. In steels, carbon strengthens the matrix via solid-solution hardening and forms carbides that impede grain boundary sliding and cavity nucleation~\cite{maiyaElevatedtemperatureLowcycleFatigue1977, leeEffectThermalAging1993, minCorrelationCharacteristicsGrain2003, shankarLowCycleFatigue2010, hongImprovementCreepfatigueLife2003, kangEffectsRecrystallizationAnnealing2010, liuEngineeringMetalcarbideHydrogen2024, chenDirectObservationIndividual2017, takahashiOriginHydrogenTrapping2018}. Boron, though typically added in trace amounts, exhibits strong grain boundary affinity, enhancing cohesion, transformation control, and hydrogen resistance~\cite{banerjiBoronSteelProceedings1980, kimEffectBoronStructure1990, cameronSolubilityBoronIron1986, heGrainBoundarySegregation1989, garlippAusteniteDecompositionCMn2001, darosaGrainboundarySegregationBoron2020, liSegregationBoronPrior2015, hachetSegregationPriorAustenite2024}. These behaviors illustrate how local chemistry and coordination at interfaces can dictate macroscopic performance under creep, fatigue, and embrittlement. In high-entropy alloys (HEAs), carbon enhances strength, twinning activity, and transformation behavior in both FCC and BCC systems~\cite{liInterstitialAtomsEnable2017, wangEffectInterstitialCarbon2016, chenEffectContentMicrostructure2018, guoEffectsCarbonMicrostructures2019}. Meanwhile boron improves grain boundary cohesion at ppm levels and promotes boride precipitation, dynamic recrystallization, and eutectic structures~\cite{leeEffectBoronCorrosion2007, zhangEvolutionMicrostructureProperties2016, houEffectsBoronContent2019, seolBoronDopedUltrastrong2018a, jiaBoronMicroalloyingHightemperature2024, wangMicrostructuralEvolutionMechanical2024, pangSimultaneousEnhancementStrength2021, zhangSignificantImprovementWear2023, sonFacileStrengtheningMethod2022, tuCharacterizationDeformationSubstructure2022}. These findings highlight the need to resolve how local coordination environments modulate the thermodynamic roles of light interstitials in chemically complex systems.

Ni-based superalloys provide an ideal platform to address this question. Boron was among the first interstitials introduced to improve grain boundary cohesion and suppress intergranular fracture, even at levels below 50~ppm~\cite{simsSuperalloysIIHighTemperature1987, garosshenEffectsZrStructure1987}. In cast systems such as IN-100 and Udimet 500, boron improved creep life and ductility, while modern strategies leverage its effects through co-segregation, additive manufacturing, and boride stabilization~\cite{bashirEffectInterstitialContent1993, chenEnhancingSulfurEmbrittlement2025, gongMicrostructuralEvolutionMechanical2023, tianSynergisticEffectsBoron2024a, tekogluSuperiorHightemperatureMechanical2024c, tekogluMetalMatrixComposite2024b, zhouHotDeformationBehavior2024, zhangInfluenceTiB2Content2022}. Carbon, in contrast, has traditionally been employed as a carbide former, precipitating as MC and M\textsubscript{23}C\textsubscript{6} carbides that impede grain boundary sliding and dislocation motion~\cite{simsSuperalloysIIHighTemperature1987, reedSuperalloysFundamentalsApplications2006}. While finely distributed carbides enhance creep life, coarse or continuous films degrade ductility~\cite{garosshenEffectsZrStructure1987}. Recent studies have expanded carbon’s role into fatigue resistance, deformation control, and in situ carbide formation during additive processing~\cite{wangInsightLowCycle2023, zhangSynergyPhaseMC2024, liInfluenceCarbidesPores2024, tekogluStrengtheningAdditivelyManufactured2023, minReactioninducedNanosizedTiC2024, ozouakiBehaviorsTiCContaining2023, hanLaserPowderBed2021}. These results underscore the contrast between carbon’s rigid bonding and phase-forming tendencies versus boron’s more chemically sensitive and structurally flexible behavior.

Classical segregation models such as McLean’s isotherm~\cite{mcleanGrainBoundariesMetals1957} treat segregation as a scalar energy difference between bulk and interface sites. While conceptually useful, this simplification fails to capture the chemical and structural heterogeneity of real interfaces in chemically complex alloys. Recent work has addressed this complexity through spectrum-based frameworks that quantify a distribution of segregation energies~\cite{wagihSpectrumGrainBoundary2019, wagihLearningGrainBoundarySegregation2022}. They reveal that grain boundaries comprise a range of atomic environments with distinct affinities for solutes, and that segregation behavior is best described statistically. Here, we extend the spectral segregation concept to interstitial solutes, introducing a framework that maps segregation energy distributions as a function of local chemical coordination. We focus on Cr substitution in a model Ni-based alloy because Cr is a principal constituent of most Ni-based superalloys, is known to influence grain boundary chemistry, and readily forms stable borides and carbides at grain boundaries~\cite{bashirEffectInterstitialContent1993, gongMicrostructuralEvolutionMechanical2023, tianSynergisticEffectsBoron2024a, tekogluSuperiorHightemperatureMechanical2024c, tekogluMetalMatrixComposite2024b, zhangSynergyPhaseMC2024, liInfluenceCarbidesPores2024, kontisEffectBoronGrain2016a, kontisRoleBoronImproving2017a}. By systematically varying the number of Cr atoms surrounding a boron or carbon atom, we resolve how Cr coordination modulates their segregation energy at a grain boundary and free surface.

\begin{figure}[H]
    \centering
    \includegraphics[width=\linewidth]{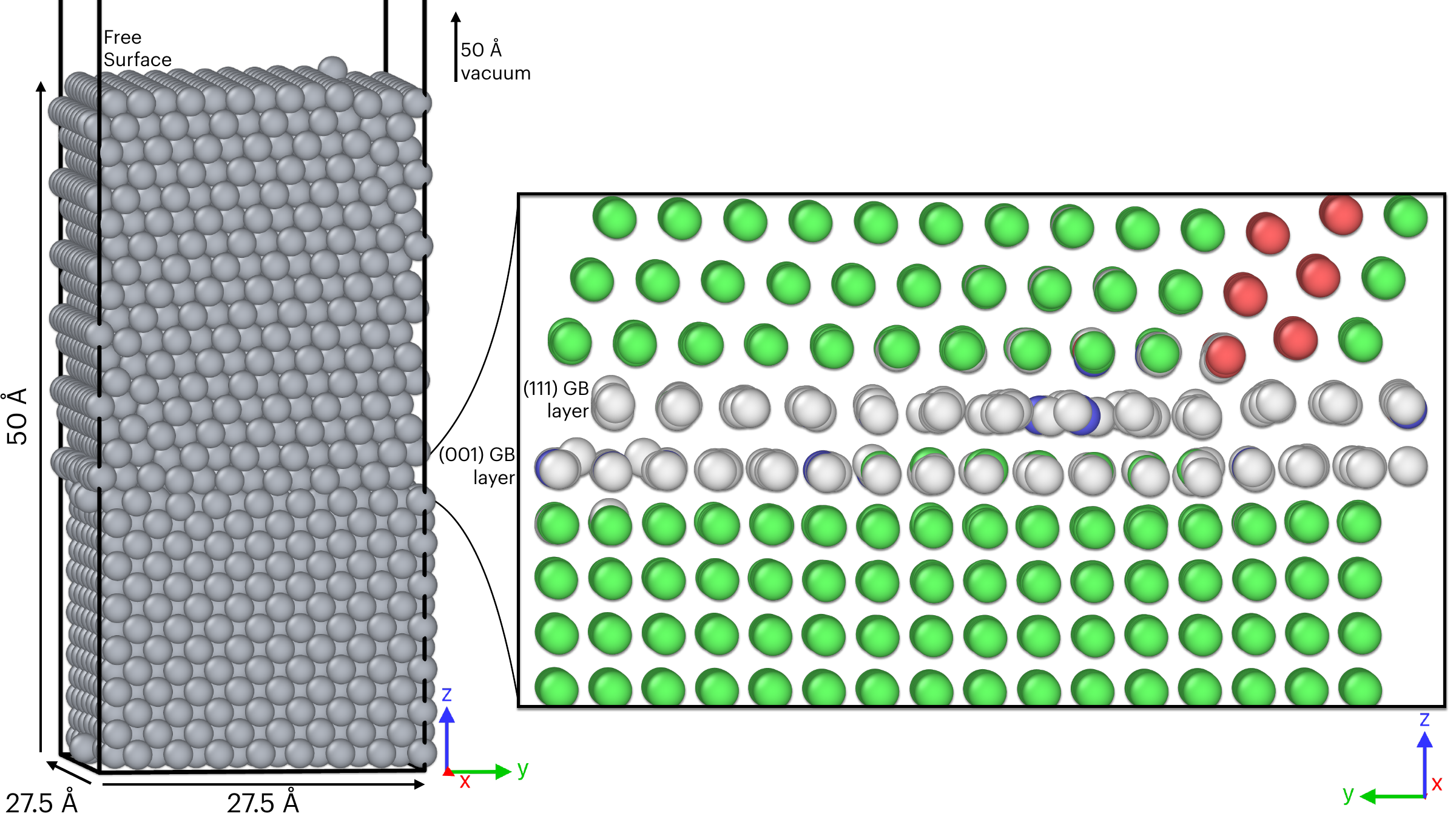}
    \caption{Simulation cell used for spectral segregation analysis. The full bicrystal is shown on the left, with a magnified view of the grain boundary (GB) on the right. Atoms are colored according to OVITO’s common neighbor analysis \cite{stukowskiVisualizationAnalysisAtomistic2009a}: green indicates FCC-coordinated atoms, white indicates ``other'' coordination, blue indicates BCC, and red indicates HCP. The GB region is dominated by non-FCC-coordinated atoms.}
    \label{fig:simcells}
\end{figure}

The grain boundary (GB) structure was constructed from pure Ni (Fm$\bar{3}$m) using Atomsk~\cite{hirelAtomskToolManipulating2015}. The top grain and bottom grain had crystallographic orientations such that the $\hat{z}$ axis was aligned perpendicular to the (111) and (001) surfaces, respectively (Fig.~\ref{fig:simcells}). Relaxation of the GB structure was performed using vacancy insertion and molecular dynamics annealing, followed by conjugate gradient minimization. Spectral segregation energies were calculated by inserting a single boron or carbon interstitial at sites spanning the bulk, GB, and free surface regions. For each interstitial site, we systematically modified the local chemical environment by progressively substituting Ni atoms in the first-nearest neighbor coordination shell of the light interstitial with Cr atoms, one at a time, until the shell was predominately Cr-occupied. This procedure ensures that each increase in $n$ replaces a direct interstitial--Ni bond with an interstitial--Cr bond, enabling a controlled assessment of how local Cr coordination affects the energetics and structure of the interstitial site. Total energies were recorded at each coordination level to generate spatially resolved segregation spectra. The segregation energy at each site was defined as:
\begin{equation}\label{eq:eseg}
E_\mathrm{seg}(n) = E_\mathrm{interface}(n) - \langle E_\mathrm{bulk}^{(n)} \rangle,
\end{equation}
where $E_\mathrm{interface}(n)$ is the total energy of the Cr-decorated interstitial configuration at either the GB or free surface and $\langle E_\mathrm{bulk}^{(n)} \rangle$ is the mean bulk reference energy at the same Cr coordination level $n$. This definition measures the relative stability of a given interstitial--Cr motif at the interface compared to the bulk, not the intrinsic binding energy between the interstitial and Cr. A negative $E_\mathrm{seg}$ indicates that the interface stabilizes that specific motif more than the bulk. All calculations employed the Matlantis implementation of the Preferred Potential (PFP)~\cite{takamotoUniversalNeuralNetwork2022,Matlantis} with LAMMPS~\cite{thompsonLAMMPSFlexibleSimulation2022a}. Full methodological details are provided in the Supplemental Materials.

\begin{figure}[H]
    \centering
    \includegraphics[width=\linewidth]{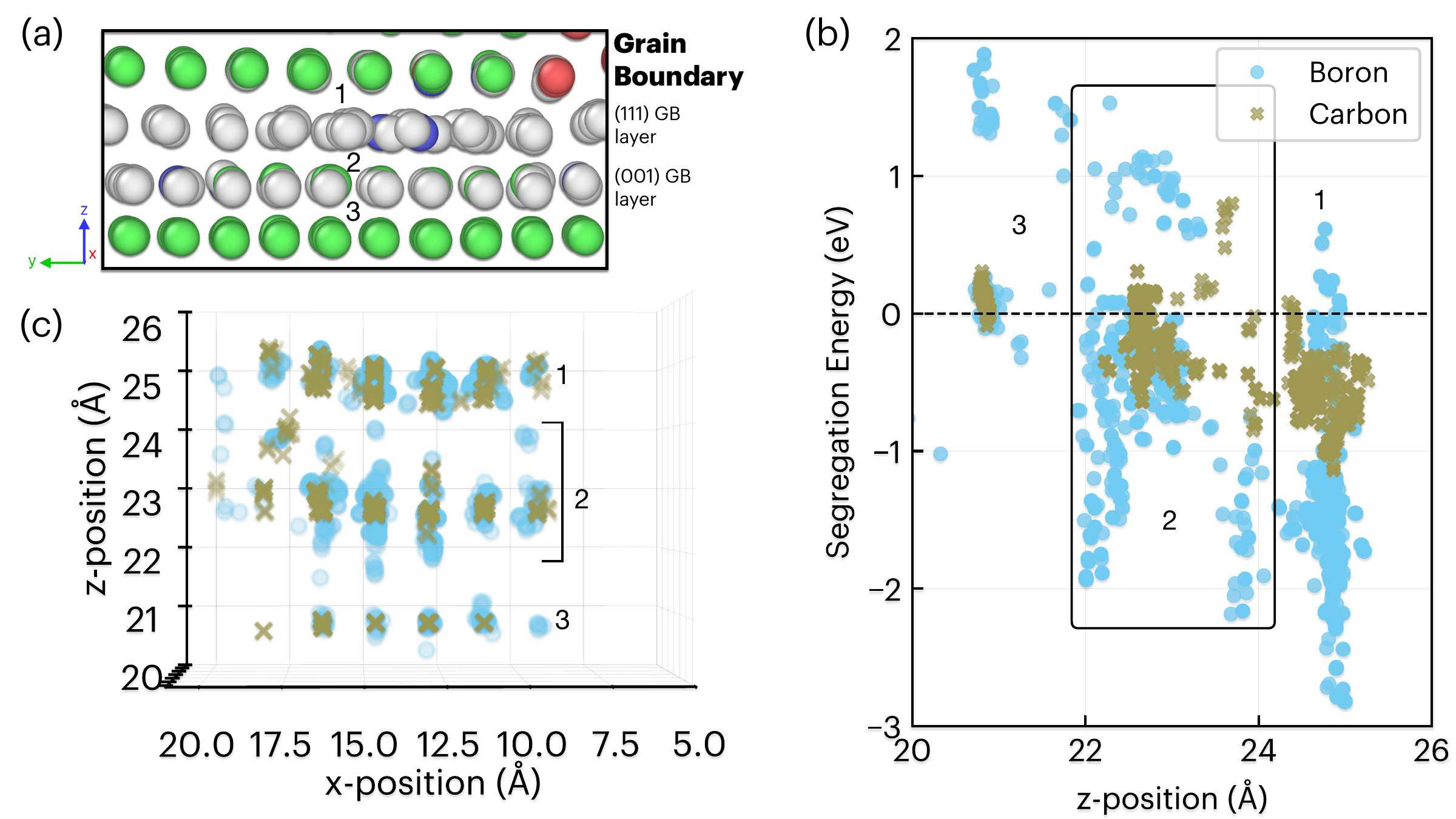}
    \includegraphics[width=\linewidth]{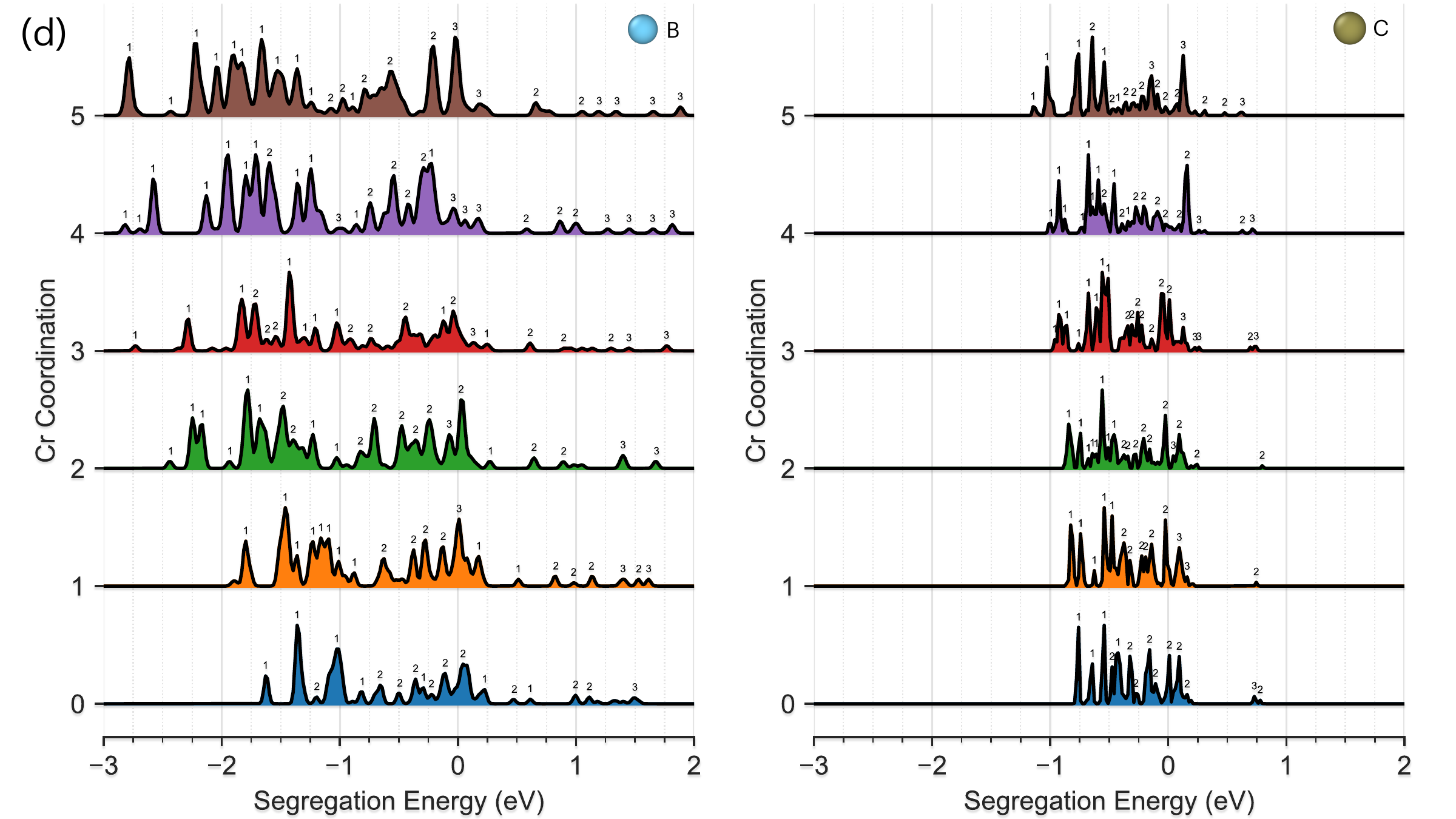}
    \caption{(a) Magnified rendering of the GB structure, with key atomic planes labeled 1, 2, and 3; these correspond to the zones marked throughout this figure. Atoms are colored according to OVITO’s common neighbor analysis: green indicates FCC-coordinated atoms, white indicates ``other'' coordination, blue indicates BCC, and red indicates HCP. The GB region is dominated by non-FCC-coordinated atoms. (b) Segregation energy for boron (blue) and carbon (gold) as a function of $z$-coordinate, compiled across all Cr decoration levels ($n = 0-5$), showing the position of the interstitial relative to the GB. (c) Three-dimensional scatter plot of boron and carbon positions from the sampling dataset. (d) Segregation energy spectra for boron (left) and carbon (right) at the GB, with each curve corresponding to a distinct number of Cr atoms in the interstitial’s first-nearest-neighbor shell, highlighting the dependence of segregation energetics on local chemistry. For clarity, the region associated with each peak is annotated above it.}
    \label{fig:gb}
\end{figure}

The $E_{\mathrm{seg}}$ data for boron and carbon at the GB are compiled in Fig.~\ref{fig:gb}. Key regions within the GB are labeled 1, 2, and 3, linking specific spectral features to their spatial origin. The raw $E_{\mathrm{seg}}$ values for $n = 0$--$5$ Cr neighbors are shown in Fig.~\ref{fig:gb}b as a function of the interstitial’s $z$-position, providing a spatially resolved view of the energetic landscape. Boron (blue) samples a broader range of GB sites than carbon (gold), including low-energy positions offset from high-symmetry planes where undercoordination and free volume enable favorable relaxation. These excursions, evident in Fig.~\ref{fig:gb}b--c, yield pronounced minima reaching $E_{\mathrm{seg}} \leq -2.0$~eV in region~1, indicating a strong driving force for segregation into specific interfacial environments. Boron’s spectrum is also more volatile: while deeply negative values are found in region~1, regions~2 and 3 exhibit highly unfavorable values exceeding $+1.0$~eV. Increasing Cr content in the first coordination shell systematically shifts boron’s spectrum toward more negative values in region~1, consistent with the formation of favorable Cr--B coordination motifs at the GB (Fig.~\ref{fig:gb}d, progressing upward). In contrast, Cr addition destabilizes boron in region~3, mirroring its bulk behavior. The breadth and variability of boron’s spectrum reflect its larger atomic radius compared to carbon~\cite{goldschmidtInterstitialAlloys1967}, which amplifies the bond-length mismatch between Ni--B and Cr--B. In the structurally compliant GB environment, these longer Cr--B bonds are accommodated through larger positional relaxations, enabling boron to stabilize a wide variety of distinct local environments.

Carbon’s $E_{\mathrm{seg}}$ distribution is markedly narrower (Fig.~\ref{fig:gb}b,d), reflecting strong site selectivity and a preference for near-perfect octahedral coordination. Its lowest energies occur in region~1, with weaker stabilization in regions~2 and 3. Across the GB region, carbon’s energies are tightly clustered, showing little of the large positive or negative excursions observed for boron. Increasing Cr content produces only a modest downward shift in region~1 and little change in regions~2 and 3, suggesting that any energetic benefit from Cr-rich environments is quickly saturated regardless of carbon’s geometric environment. This limited adaptability stems from carbon’s smaller atomic radius~\cite{goldschmidtInterstitialAlloys1967} and the correspondingly smaller bond-length mismatch between Ni--C and Cr--C, which minimize geometric relaxation even when Cr occupies first-neighbor positions. As a result, carbon shows little variability in its positional data (Fig.~\ref{fig:gb}b,c) in regions~1 and 2, suggesting it cannot access the diverse bonding environments that stabilize boron at the GB. Boron’s broad, widely distributed spectrum indicates a rugged PES with many local minima and maxima, reflecting its ability to explore and stabilize diverse coordination environments at the GB. In contrast, carbon’s narrow spectrum corresponds to a smoother PES dominated by a single preferred coordination state, consistent with its rigid site selectivity and limited structural adaptability.

\begin{figure}[H]
    \centering
    \includegraphics[width=\linewidth]{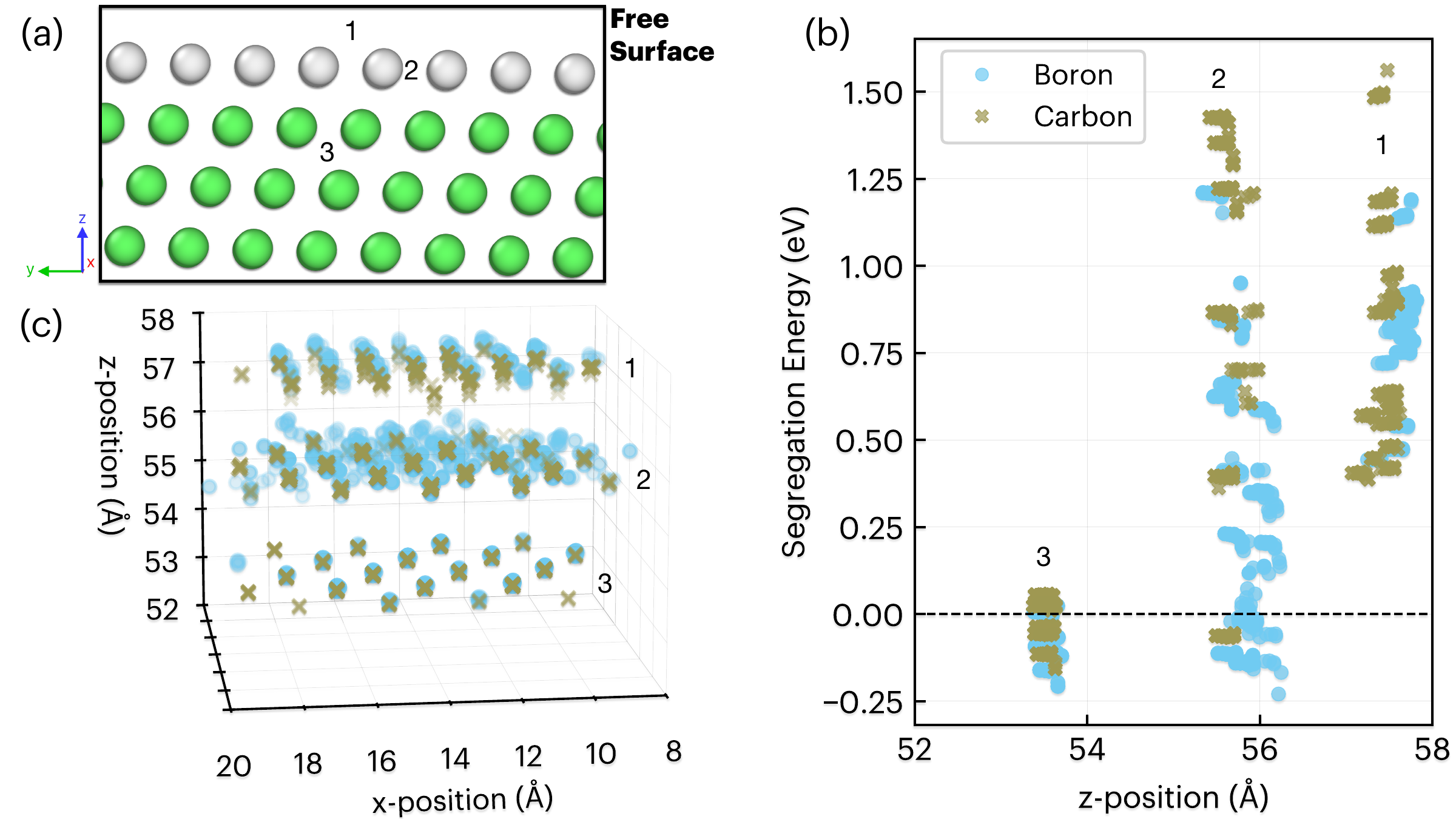}
    \includegraphics[width=\linewidth]{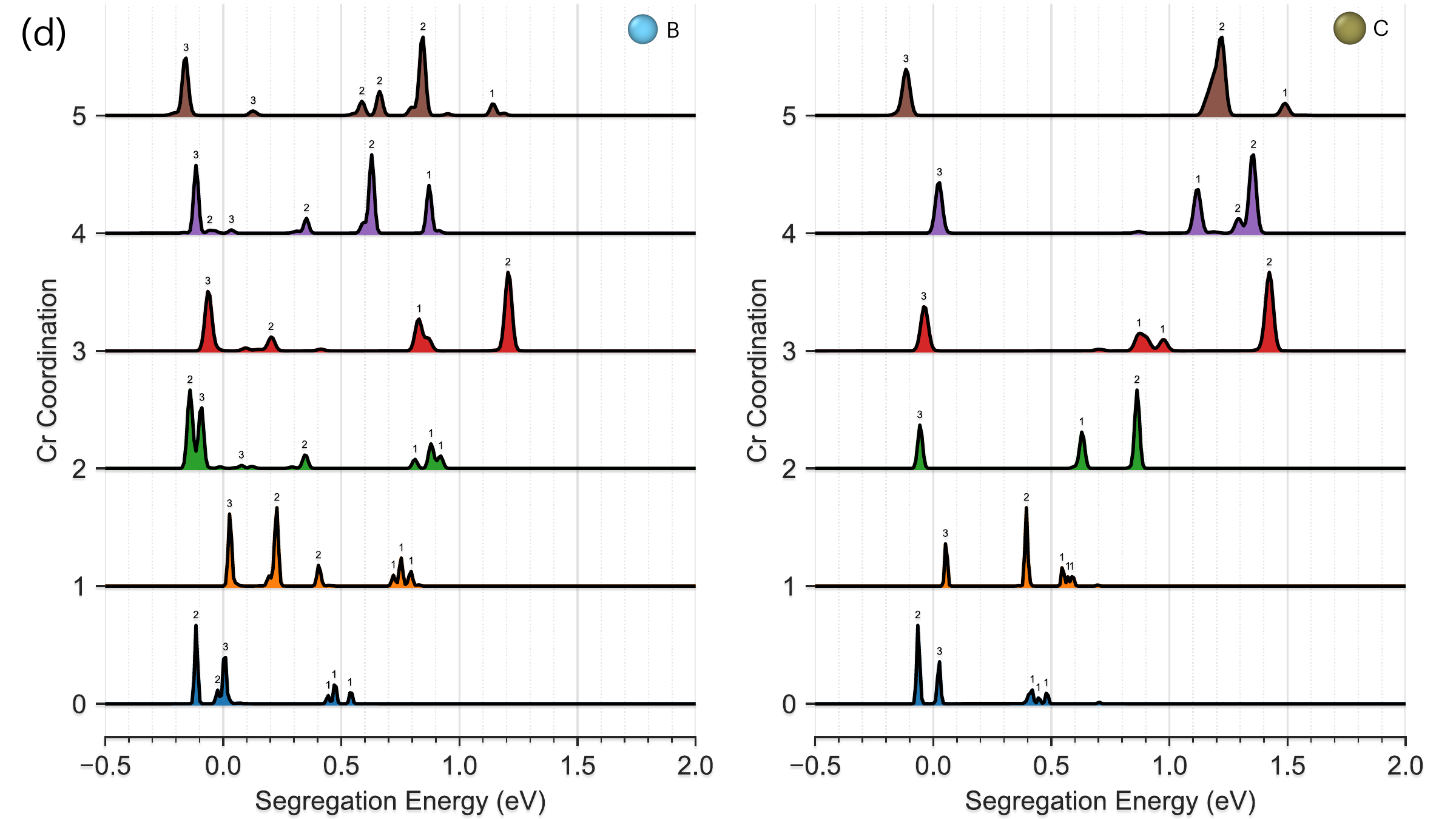}
    \caption{(a) Magnified rendering of the free surface region, with key atomic planes labeled 1, 2, and 3; these correspond to the zones marked throughout this figure. Atoms are colored according to OVITO’s common neighbor analysis: green indicates FCC-coordinated atoms and white indicates ``other'' coordination. (b) Segregation energy for boron (blue) and carbon (gold) as a function of $z$-coordinate, compiled across all Cr decoration levels ($n = 0-5$), showing the position of the interstitial relative to the free surface. (c) Three-dimensional scatter plot of boron and carbon positions from the sampling dataset. (d) Segregation energy spectra for boron (left) and carbon (right) at the free surface, with each curve corresponding to a distinct number of Cr atoms in the interstitial’s first-nearest-neighbor shell. For clarity, the region associated with each peak is annotated above it.}
    \label{fig:surface}
\end{figure}

The $E_{\mathrm{seg}}$ data for boron and carbon at the free surface are compiled in Fig.~\ref{fig:surface}. Key regions are labeled 1, 2, and 3, with region~1 corresponding to adatom sites atop the surface. Moving into subsurface voids (regions~2 and 3), the $E_{\mathrm{seg}}$ distribution collapses toward zero or moderately negative values (Fig.~\ref{fig:surface}b,d), reflecting increased coordination and partial screening of the vacuum interface. In region~2, boron samples a broader energetic range than carbon and accesses substantially more negative states, despite this region being embedded within the surface atomic plane and not yet fully coordinated as in region~3. The positional data in Fig.~\ref{fig:surface}c corroborate this, showing many configurations sampled by boron that are inaccessible to carbon. These results suggest that partially coordinated surface sites can provide favorable bonding environments for boron, whereas carbon remains constrained to well-ordered, high-symmetry configurations. Unlike the broad, multi-peaked spectra at the GB, free-surface spectra cluster into a few discrete states, reflecting the reduced geometric complexity of the surface environment.

At the free surface, $E_{\mathrm{seg}}$ for both boron and carbon increases with Cr content in the first-nearest-neighbor shell in regions~1 and 2, indicating destabilization relative to the subsurface region~3. The effect is stronger for carbon, reflecting truncated Cr coordination that disrupts the extended Cr--C bonding network required for stability. Cr--C bonding is highly coordination-dependent: stable carbides such as Cr\textsubscript{3}C and Cr\textsubscript{7}C\textsubscript{3} exhibit well-hybridized C\textsubscript{p}--Cr\textsubscript{d} states, whereas Cr\textsubscript{23}C\textsubscript{6} shows broader states and partial Cr\textsubscript{d} occupation at the Fermi level~\cite{liElectronicMechanicalProperties2011}. In contrast, Ni-rich carbides (e.g., Ni\textsubscript{3}C) display bonding states fully below the Fermi level~\cite{hubaMonitoringFormationCarbide2014, jainCommentaryMaterialsProject2013a}, consistent with complete bond saturation; similar trends are found in Ni--B systems~\cite{jainCommentaryMaterialsProject2013a}. \textit{Ab initio} calculations further show that clean Ni~(111) surfaces maintain a filled $d$-band, whereas Cr substitution introduces partially filled, reactive $d$-states that are especially destabilizing at undercoordinated sites~\cite{mittendorferStructuralElectronicMagnetic1999}. This electronic mismatch explains the observed $E_{\mathrm{seg}}$ trends and underpins the thermodynamic preference for boron and carbon to segregate to GBs rather than internal free surfaces without adsorbates.

Considering both GB and free-surface segregation, these results indicate a thermodynamic driving force that pulls interstitials away from internal surfaces and toward grain boundaries. At the surface, Cr-rich coordination destabilizes both boron and carbon, raising $E_{\mathrm{seg}}$ and limiting the number of favorable configurations. In contrast, the same Cr environments at the GB permit greater bonding flexibility, strain accommodation, and charge redistribution. This is especially true for boron, which stabilizes a wide variety of sites, including low-symmetry positions inaccessible to carbon. This inversion reflects the shift from electronically frustrated, undercoordinated surface sites to chemically accommodating GB environments. The resulting energetic contrast establishes a natural segregation gradient, driving interstitials from higher-energy internal surfaces (e.g., voids, microcracks) toward lower-energy GB sites. Importantly, this intrinsic behavior differs from that at oxidized surfaces, where boron and Cr co-segregate in the presence of oxygen~\cite{dolezalAtomisticMechanismsOxidation2025}. In the absence of adsorbates, segregation is governed by local metallic coordination and bonding flexibility rather than oxide-driven stabilization.

\begin{figure}[H]
    \centering
    \includegraphics[width=\linewidth]{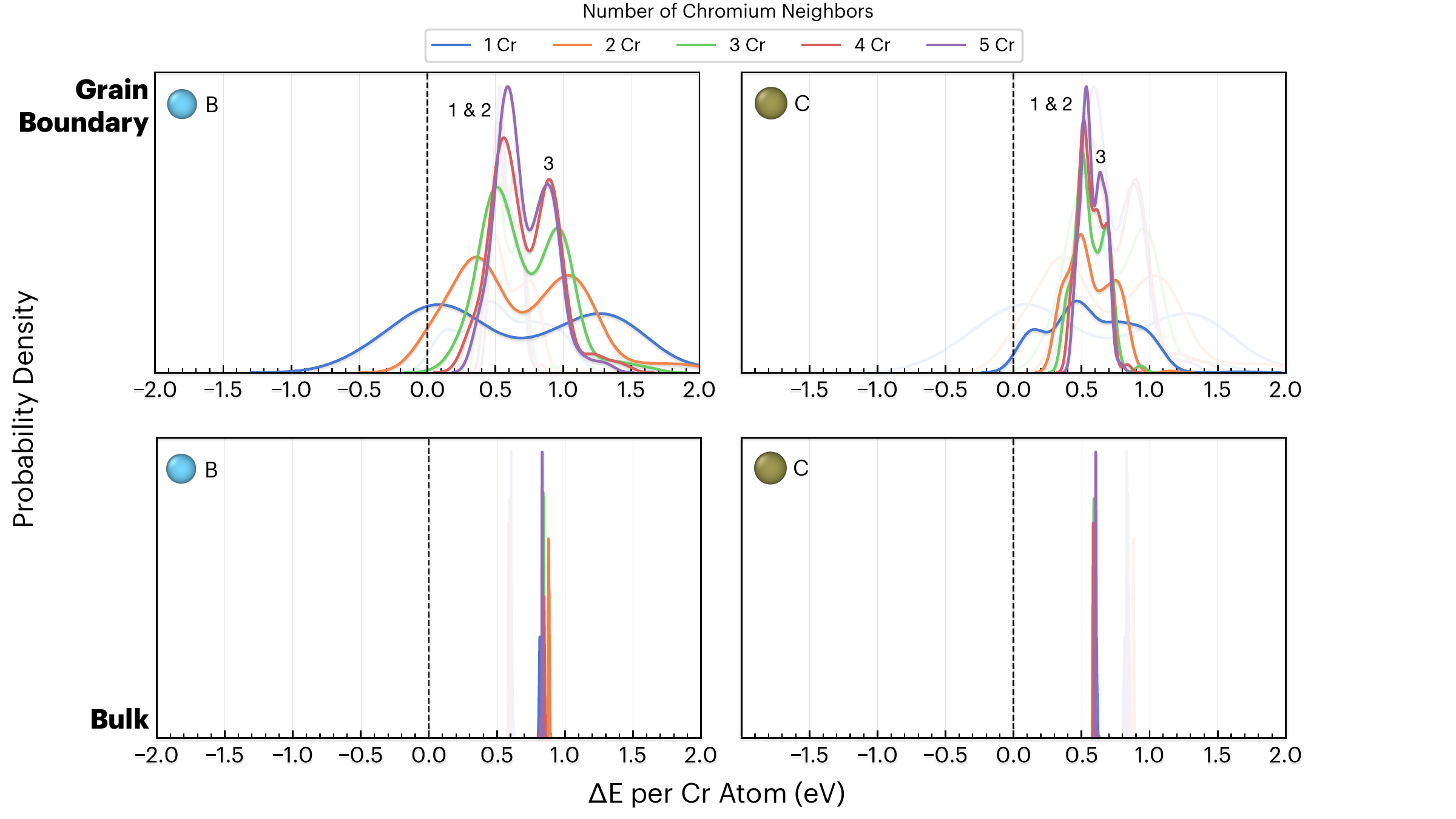}
    \caption{Change in total energy for boron (left) and carbon (right) as a function of Cr coordination, shown for grain boundary sites (top) and bulk sites (bottom). For direct comparison, carbon curves are overlaid with reduced opacity on boron panels, and vice versa.}
    \label{fig:energy}
\end{figure}

The bottom row of Fig.~\ref{fig:energy} shows that adding Cr atoms to the first coordination shell of a light interstitial imposes a substantial energy penalty in the bulk. The total energy rises by approximately 0.6~eV per Cr atom around carbon and 0.84~eV per Cr around boron. These positive values reflect the chemical mismatch in the Ni-rich bulk, where Cr--C and Cr--B bonding is electronically frustrated. At the GB this penalty is partially mitigated (top row of Fig.~\ref{fig:energy}). In regions 1 and 2, boron’s energy cost decreases towards 0.6~eV per Cr atom, whereas carbon shows little improvement. This divergence indicates that Cr-rich environments at the GB are more accommodating for boron than carbon. Although Cr addition remains energetically costly overall, the relative reduction for boron aligns with its broader, more negative segregation spectrum and its ability to stabilize a wider variety of GB configurations.

It is important to emphasize that $E_{\mathrm{seg}}$ (Eq.~\ref{eq:eseg}) measures the relative favorability of a Cr--interstitial configuration at the GB compared to the bulk, not its absolute formation energy. Even when Cr addition near boron is costly in both environments, the GB stabilizes the system through relaxation, charge redistribution, and strain relief unavailable in the bulk lattice. A negative $E_{\mathrm{seg}}$ thus indicates that the GB converts a locally frustrated configuration into a globally favorable one. This effect is most pronounced for Cr--B interactions in region~1 of the GB, where $E_{\mathrm{seg}}$ becomes increasingly negative with Cr content despite local penalties, reflecting boron’s ability to exploit undercoordinated and asymmetric sites (Fig.~\ref{fig:gb}). In contrast, Cr--C interactions remain costly and structurally constrained, consistent with carbon’s narrower $E_{\mathrm{seg}}$ distribution and limited site diversity. These results show that GBs act not merely as sinks for solutes but as chemically transformative environments that stabilize bonding motifs inaccessible in the bulk. This interpretation aligns with our previous finding that B--Cr co-segregation to a Ni--Cr GB reduces $\gamma_{\mathrm{GB}}$~\cite{dolezalSegregationOrderingLight2025}, underscoring the role of such motifs in GB stabilization.

While the present study focuses on a low-energy GB for its structural simplicity and suitability for systematic spectral sampling, this boundary type represents only a minority fraction in polycrystalline Ni-based superalloys, where random high-angle boundaries dominate the network~\cite{schuhUniversalFeaturesGrain2005, rohrerComparingCalculatedMeasured2010}. The spectral segregation framework introduced here is not limited to a specific GB character and can be applied to any GB geometry. Although absolute segregation energies will vary with GB structure, the underlying picture that emerges, a broad and rugged spectrum of energetics rather than a single-valued quantity, is expected to persist. The very complexity of the spectra highlights the chemical and structural heterogeneity of GBs, where interstitials encounter a diverse landscape of favorable and unfavorable sites. Notably, Cr-based borides and carbides are among the most frequently observed GB-anchored precipitates in Ni-based superalloys~\cite{bashirEffectInterstitialContent1993, gongMicrostructuralEvolutionMechanical2023, tianSynergisticEffectsBoron2024a, tekogluSuperiorHightemperatureMechanical2024c, tekogluMetalMatrixComposite2024b, zhangSynergyPhaseMC2024, liInfluenceCarbidesPores2024, kontisEffectBoronGrain2016a, kontisRoleBoronImproving2017a}, underscoring the practical relevance of Cr--B and Cr--C interactions to GB chemistry.

This work reframes light interstitial segregation as a distributional response shaped by local chemistry and interfacial structure. Extending the spectral approach originally developed for substitutional solutes, we show that carbon remains confined to narrow, sharply peaked states with minimal displacement, whereas boron explores a rugged energetic landscape with greater positional flexibility. The contrast between destabilizing Cr environments at free surfaces and stabilizing Cr environments at GBs establishes a natural segregation gradient that drives interstitials toward boundaries. The complexity of the spectra reflects the true heterogeneity of interfacial PES landscapes rather than noise. Spectral sampling provides a high signal-to-noise method for resolving these environments and can be coupled with Monte Carlo, kinetic, or machine learning approaches to generate site-specific, finite-temperature segregation models. Such datasets offer a pathway to alloy design strategies that leverage interfacial chemistry to improve cohesion, creep resistance, and oxidation tolerance.

\section*{Author Contributions}
\textbf{T.D.D} Writing - original draft, writing - review \& editing, visualization, validation, software, methodology, investigation, formal analysis, data curation. \textbf{R.F.} Project administration and supervision, writing - review \& editing. \textbf{J. Li} Project administration and supervision, writing - review \& editing, funding acquisition. All authors contributed to the conceptualization of this project.

\section*{Acknowledgments}
J. Li acknowledges support from National Science Foundation, USA CMMI-1922206 and DMR-1923976.

\bibliographystyle{ieeetr}
%\bibliography{main}

\end{document}